\documentclass[a4paper,fleqn,usenatbib]{mnras}

\usepackage[T1]{fontenc}
\usepackage{ae,aecompl}
\usepackage[utf8]{inputenc}

\usepackage{graphicx}	
\usepackage{amsmath}	
\usepackage{amssymb}	
\usepackage{lscape}
\usepackage{tikz}   
\usepackage{gensymb}
\usepackage{tikz-3dplot} 
\usepackage{import}
\usepackage{lmodern}
\RequirePackage{fix-cm}

\usepackage{graphicx}	
\usepackage{amsmath}	
\usepackage{amssymb}	
\usepackage{grffile}
\usepackage{hyperref}

\newcommand{\ex}{\mathbf{e}_{\rm x}}
\newcommand{\ey}{\mathbf{e}_{\rm y}}
\newcommand{\ez}{\mathbf{e}_{\rm z}}

\newcommand{\rlight}{r_{\rm L}}

\title[Rotating off-centred dipoles]{Impact of an off-centred dipole on neutron star binaries}

\author[J. P\'etri]{
J. P\'etri\thanks{E-mail: jerome.petri@astro.unistra.fr}
\\
Universit\'e de Strasbourg, CNRS, Observatoire astronomique de Strasbourg, UMR 7550, F-67000 Strasbourg, France.
}

\date{Accepted XXX. Received YYY; in original form ZZZ}

\pubyear{2019}

\begin{document}
\label{firstpage}
\pagerange{\pageref{firstpage}--\pageref{lastpage}}
\maketitle

\begin{abstract}
Neutron stars are strongly magnetized rotating compact objects. Therefore they also produce huge electric fields accelerating particles to ultra-relativistic energies. The simplest magnetic topology is a dipole traditionally located at the stellar centre. In this paper, we reinvestigate the consequences of an off-centred rotating magnetic dipole, showing accurate magnetic field line geometries, the associated spin-down luminosity as well as the corresponding electromagnetic kick and torque imprinted to the neutron star. Results are obtained by time dependent numerical simulations of Maxwell equations in vacuum using pseudo-spectral methods. We compare our results to known analytical expressions available to lowest order in the parameter $\epsilon = d/R$, where $d$ is the displacement of the dipole from the stellar centre and $R$ the neutron star radius. We found good agreement between our numerical computations and our analytical approximations even for well off-centred dipoles having large displacements with a sizeable fraction of the radius, i.e.~$\epsilon \lesssim 1$. An explanation for binary neutron star eccentricity distribution functions is given with an emphasize on highly eccentric systems as an alternative scenario to traditional binary formation.
\end{abstract}




\section{Introduction}

Magnetic field and rotation play an important role in the life and evolution of stars. This is especially true for magnetized compact objects like white dwarfs and neutron stars. There the magnetic field is often assumed to be dipolar, which corresponds to the lowest order expansion of any magnetic field geometry not possessing magnetic monopoles. Moreover, for spherical stars, the dipolar magnetic moment is located exactly at the centre of the star. The main reason for this assumption is simplicity. However, such approximation is hardly supported by any physical explanation. Actually the probability to find such a coincidence seems rather weak. It is conceivable that currents in the stellar interior lead to field structures not centred and also not strictly dipolar. For strongly magnetized stars such as neutron stars, this asymmetry in the topology could have observable consequences. \cite{harrison_acceleration_1975} already mentioned the possibility to explain the high kick velocity above 100~km/s induced by the electromagnetic force imprinted to the newly born star. Some refinements were soon brought by \cite{tademaru_acceleration_1976} who added polarization effects and thus contributions from an additional electric field component.

In the case of a centred dipole, the torque applied on a neutron star in vacuum always predicts an alignment as explained by \cite{davis_magnetic-dipole_1970} and \cite{goldreich_neutron_1970}. This fact is also discussed by \cite{philippov_time_2014}. Moreover, this electromagnetic torque has been computed following several expressions, either using the electromagnetic stress-energy tensor or the Lorentz force acting inside and on the surface of the neutron star. Both approaches agree on the regular electromagnetic torque but not on the anomalous torque. A single analytical formula has been found for this anomalous torque but the scaling factor depends on the expression used to compute the torque, from the stress-energy tensor or directly from the Lorentz force. Also the electric field contribution, sometimes omitted, should be included, see for instance \cite{beskin_anomalous_2014} for a critical review of these discrepancies. These complications did not prevent \cite{good_electromagnetic_1985} from considering contributions from a quadrupole component to the total torque. 

Decentred dipoles furnish simple explanations for the observations of planetary magnetic fields. Indeed, \cite{komesaroff_centred_1976} already recognized that a decentred dipole better fits the Pioneer data about Jupiter than a centred dipole. It was used to explain the asymmetries in the radio emission by minimizing the quadrupolar term. \cite{landstreet_measurement_1980} presented a review about magnetic fields in non degenerated stars and concluded that a decentred oblique dipole can reasonably fit the magnetic topology. Hints for off-centred dipoles or dipole plus quadrupole fields are given by \cite{putney_off-centered_1995} from polarization observations of white dwarfs. In our solar system some interesting results have been found for the outer planets. Indeed, inclination angles and displacements have been constrained for instance for Uranus with an inclination of $60\degr$ and an off-centring of $d = 0.3\,R_{\rm U}$ \citep{ness_magnetic_1986} and Neptune with an inclination of $47\degr$ and an off-centring of $d = 0.55\,R_{\rm N}$ \citep{ness_magnetic_1989} where $R_{\rm U}$ and $R_{\rm N}$ are Uranus and Neptune radius respectively. However, distinguishing an off-centred dipole from a combination of multipoles seems difficult on the base only of observations \citep{martin_magnetic_1984}.

Deviation from a pure and centred rotating dipole is also an attractive assumption to get more insight into neutron star and pulsar physics. Indeed, a simple prescription for a distorted dipolar field was proposed by \cite{harding_pulsar_2011} to enhance the pair production rate above polar caps. Non dipolar fields have also been suggested to explain the anomalous braking index of pulsars as discussed by \cite{barsukov_influence_2010}. With the wealth of new and very accurate multi-wavelength observations of pulsed emission in pulsars including polarization, time is ripe to go further than the almost exclusively used centred dipole to sharpen our view on pulsar magnetospheres. An off-centred rotating dipole anchored in a perfectly conducting sphere with finite radius leads very naturally to multipolar components to any order with weighting coefficients solely related to the geometry of the dipole. We use this assumption as a starting point for the justification of multipolar components. 

In this paper we compute accurate numerical solutions for the electromagnetic field in vacuum outside an off-centred rotating dipole, valid for any displacement~$d$ such that $\epsilon = d/R<1$, where $d$ is the distance from the centre of the neutron star and $R$ its radius. In Sec.~\ref{sec:Modele}, we recall the basic model of an off-centred rotating dipole. Then some examples of field lines are presented in Sec.~\ref{sec:LigneChamp}. Next, in Sec.~\ref{sec:Luminosite} we compute the spin-down luminosity expected from such a system and compare it with previous works. The same comparison is done for the associated electromagnetic kick which is estimated in Sec.~\ref{sec:Kick} and the electromagnetic torque in Sec.~\ref{sec:Couple}. Implications for binaries containing neutron stars are highlighted in Sec.~\ref{sec:Binaire}. Conclusions are drawn in Sec.~\ref{sec:Conclusion}.

\section{Rotating off-centred dipole}
\label{sec:Modele}

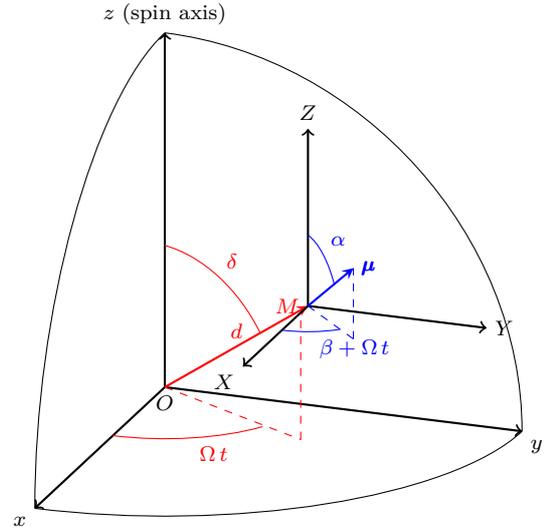
\begin{figure}
	\centering
	
	
	\tdplotsetmaincoords{70}{110}
	
	\pgfmathsetmacro{\rvec}{.6}
	\pgfmathsetmacro{\thetavec}{55}
	\pgfmathsetmacro{\phivec}{70}
	
	\begin{tikzpicture}[scale=5,tdplot_main_coords]
	
	\coordinate (O) at (0,0,0);
	
	\tdplotsetcoord{P}{\rvec}{\thetavec}{\phivec}
	
	\tdplotsetcoord{A}{1}{60}{0}
	
	\tdplotsetcoord{B}{1.5}{60}{0}
	
	
	\draw[thick,->] (0,0,0) node[below] {$O$} -- (1,0,0) node[anchor=north east]{$x$};
	\draw[thick,->] (0,0,0) -- (0,1,0) node[anchor=north west]{$y$};
	\draw[thick,->] (0,0,0) -- (0,0,1) node[anchor=south]{$z$ (spin axis)};
	
	\tdplotsetrotatedcoords{0}{0}{0}
	
	\draw[-stealth,thick,color=red] (O) -- (P) node [midway, above] {$d$} node [above, left] {$M$} ;
	\draw[dashed,color=red,tdplot_rotated_coords] (0,0,0) -- (0.26,0.475,0);
	\draw[dashed,color=red,tdplot_rotated_coords] (0.26,0.475,0) -- (0.26,0.475,.35);
	\tdplotdrawarc[tdplot_rotated_coords,color=red]{(0,0,0)}{0.4}{0}{60}{anchor=north west}{$\Omega\,t$}
	
	
	
	
	\tdplotsetthetaplanecoords{\phivec}
	
	\tdplotdrawarc[red, tdplot_rotated_coords]{(0,0,0)}{0.4}{0}{\thetavec}{anchor=south west}{$\delta$}
	
	\tdplotsetthetaplanecoords{0}
	
	
	
	\tdplotsetrotatedcoords{0}{0}{0}
	
	\tdplotsetrotatedcoordsorigin{(P)}
	
	\draw[thick,tdplot_rotated_coords,->] (0,0,0) -- (.5,0,0) node[anchor=north east]{$X$};
	\draw[thick,tdplot_rotated_coords,->] (0,0,0) -- (0,.5,0) node[anchor=west]{$Y$};
	\draw[thick,tdplot_rotated_coords,->] (0,0,0) -- (0,0,.5) node[anchor=south]{$Z$};
	
	
	\draw[-stealth,thick,color=blue,tdplot_rotated_coords] (0,0,0) -- (.2,.2,.2) node [right] {$\pmb{\mu}$} ;
	\draw[dashed,color=blue,tdplot_rotated_coords] (0,0,0) -- (.2,.2,0);
	\draw[dashed,color=blue,tdplot_rotated_coords] (.2,.2,0) -- (.2,.2,.2);
	
	\tdplotdrawarc[tdplot_rotated_coords,color=blue]{(0,0,0)}{0.2}{0}{45}{anchor=north west}{$\beta+\Omega\,t$}
	
	\tdplotsetrotatedthetaplanecoords{45}
	
	\tdplotdrawarc[tdplot_rotated_coords,color=blue]{(0,0,0)}{0.2}{0}{55}{anchor=south west}{$\alpha$}
	
	\begin{scope}[canvas is xy plane at z=0]
	\draw (1,0) arc (0:90:1);
	\end{scope}
	\begin{scope}[canvas is xz plane at y=0]
	\draw (1,0) arc (0:90:1);
	\end{scope}
	\begin{scope}[canvas is yz plane at x=0]
	\draw (1,0) arc (0:90:1);
	\end{scope}
	
	\end{tikzpicture}
	\caption{Geometry of an off-centred dipole showing the angles $\{\alpha, \beta, \delta\}$ and the distance~$d$. The plot corresponds to time~$t$ assuming that $\bmu$ lies in the $(xOz)$ plane at $t=0$.}
	\label{fig:Dipole}
\end{figure}

We remind the geometrical set-up for an off-centred dipole, following the notations introduced by \cite{petri_radiation_2016}. The neutron star is a perfectly conducting sphere of radius~$R$ in solid body rotation at an angular rate~$\Omega$. Its magnetic moment~$\bmu$ is located inside this sphere at a point~$M$ such that at any time~$t$ its position vector is decomposed into
\begin{equation}
\mathbf{d} = d \, (\sin \delta \, \cos(\Omega\,t) \, \ex + \sin \delta \, \sin (\Omega\,t) \, \ey + \cos \delta \, \ez )
\end{equation}
where $d$ is the distance from the stellar centre and $\delta$ the colatitude and $(\ex,\ey,\ez)$ is a cartesian orthonormal basis, Fig.~\ref{fig:Dipole}. The relevant displacement is given by the normalised parameter $\epsilon=d/R<1$. Entrainment by the star is included in the phase term $\Omega\,t$. Meanwhile, the magnetic moment~$\bmu$ points toward a direction depicted by the angles~$(\alpha,\beta)$ and given by the unit vector $\bmu = \mu \, \mathbf{m}$ such that
\begin{equation}
\mathbf{m} = \sin \alpha \, \cos (\beta+\Omega\,t) \, \ex + \sin \alpha \, \sin (\beta+\Omega\,t) \, \ey + \cos \alpha \, \ez .
\end{equation}
All important parameters are summarized in Fig.~\ref{fig:Dipole}.
Inside the star, the magnetic field is given by a static dipole such that
\begin{equation}
\label{eq:OffcenteredDipole}
\mathbf{B} = \frac{B\,R^3}{\|\mathbf{r} - \mathbf{d}\|^3} \, \left[ \frac{3\,\bmath{\mu} \cdot (\mathbf{r} - \mathbf{d})}{\|\mathbf{r} - \mathbf{d}\|^2} \, (\mathbf{r} - \mathbf{d}) - \bmath{\mu} \right]
\end{equation}
where $B$ is the surface magnetic field at the equator and $\mathbf{r}$ the position vector.

Starting from this static solution, we perform time dependent numerical simulations solving Maxwell equations in vacuum. We use the pseudo-spectral code developed and discussed in detail in \cite{petri_general-relativistic_2014}, applying it to flat spacetime. Boundary conditions at the stellar surface are given by the continuity of the radial component of the magnetic field~$B_r$ and the tangential component of the electric field where $\mathbf{E}' = \mathbf{0}$ in the corotating frame inside the star. At large distances, we enforce outgoing wave boundary conditions.

We run several sets of simulations to scan a full range of geometries, varying the angles $\alpha$, $\beta$ and $\delta$ and the displacement~$d$. We next summarized our results by first showing some magnetic field line structures, then compute the spin down luminosities and eventually the associated electromagnetic kick and torque.

\section{Field lines}
\label{sec:LigneChamp}

As a first result of a rotating off-centred dipole, we plot magnetic field lines in the equatorial plane for a perpendicular rotator with $\alpha = \delta = 90\degr$ for several values of the displacement~$d$ and angle~$\beta$. For small displacements~$d\ll R$ the topology is very similar to the centred dipole given by Deutsch solution. In all cases, a two armed spiral develops, rotating at a constant speed equal to the rotation period of the star~$\Omega$. 

\begin{figure}
	\centering
	\input{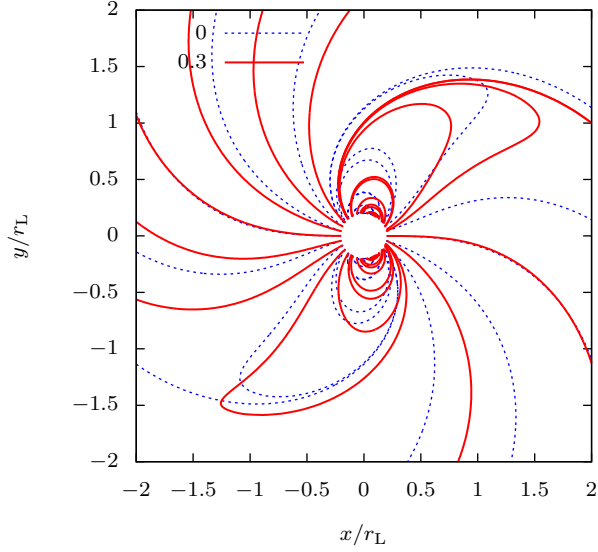}
	\caption{Magnetic field lines for an off-centred dipole with $\alpha=90\degr$, $\beta=0\degr$, $\delta=90\degr$ and $\epsilon=0.3$ (red solid line) compared to Deutsch solution $\epsilon=0$ (blue dashed line).}
	\label{fig:LigneChampB0}
\end{figure}

An example of field lines is shown for $\beta=0\degr$ and $\epsilon=0.3$ in Fig.~\ref{fig:LigneChampB0} in red solid line and compared to the centred dipole given by Deutsch solution in blue dashed line. At large distances, in the wave zone we observed a shift in phase for the position of the two-armed spiral with respect to the centred case. This shift is proportional to $R/\rlight$, therefore too weak to be detectable for realistic pulsar parameters. Again, an off-centred dipole can only reasonably be detected when examining electromagnetic activity and radiation occurring close to the surface.

\begin{figure}
	\centering
	\input{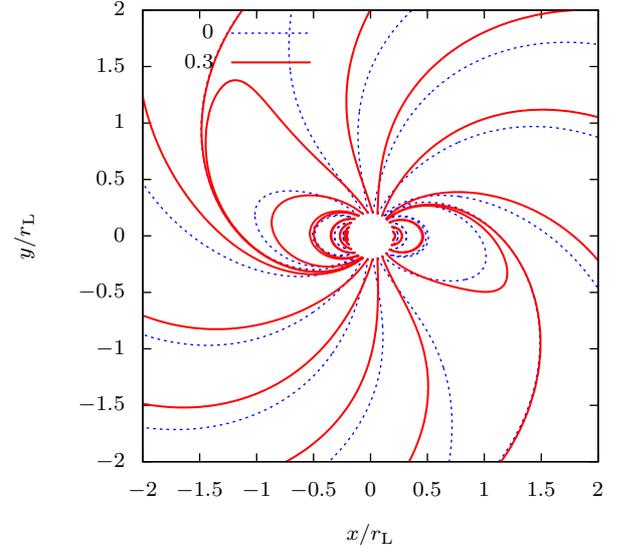}
	\caption{Magnetic field lines for an off-centred dipole with $\alpha=90\degr$, $\beta=90\degr$, $\delta=90\degr$ and $\epsilon=0.3$ (red solid line) compared to Deutsch solution $\epsilon=0$ (blue dashed line).}
	\label{fig:LigneChampB90}
\end{figure}

Another example is shown for $\beta=90\degr$ and $\epsilon=0.3$ in Fig.~\ref{fig:LigneChampB90} in red solid line and compared to the centred dipole given by Deutsch solution in blue dashed line. Here again a two-armed spiral forms, reminiscent of the centred dipole. However, in this second case, at large distance, well outside the light-cylinder, both structure are very similar, only the dipolar part propagates significantly into the wave zone. It is hard to detect an off-centred dipole by inspecting the large scale field structure because the typical two-armed spiral structures overlap. The asymmetry is only clearly visible close to the neutron star surface.

Next we diagnose quantitatively the effect of an off-centering by computing relevant physical parameters that are the spindown luminosity, the electromagnetic kick and its associated torque.

\section{Spindown luminosities}
\label{sec:Luminosite}

The spindown luminosity is an important characteristic of any rotating magnetic multipole. It determines the secular evolution of the rotational period. In \cite{petri_radiation_2016}, we showed that a finite size off-centred dipole looses angular momentum mainly through its $m=1$ (dipole and quadrupole) and $m=2$ (quadrupole) modes such that each contribution splits into
\begin{subequations}
	\label{eq:Lm1e2}
	\begin{align}
	L_{m=1} & = L_{\rm dip} \, \left[ \left( 1 - a^2 \right) \, \sin^2\alpha + \frac{24}{25} \, a^2 \, \epsilon^2 \, \cos^2\alpha \right] \\
	L_{m=2} & = \frac{48}{5} \, L_{\rm dip} \, a^2\,\epsilon^2 \, \sin^2\alpha \ .
	\end{align}
\end{subequations}
where $a=R/\rlight$ and the centred perpendicular dipole spindown is
\begin{equation}
L_{\rm dip} = \frac{8\,\upi}{3\,\mu_0\,c^3} \, \Omega^4 \, B^2 \, R^6 .
\end{equation}
For brevity, we only showed expressions~(\ref{eq:Lm1e2}) valid for $\delta=90\degr$ although it is possible to give (lengthy) expressions for any angle~$\delta$. Note that the quadrupolar contribution to the spindown is of the same order of magnitude as the perturbation in the dipole spindown. Both terms scale as $a^2\,\epsilon^2$ but show a different dependence with respect to the inclination angle~$\alpha$, the $m=1$ mode adds a $\cos^2\alpha$ contribution whereas the $m=2$ mode adds a $\sin^2\alpha$ contribution. However, both corrections are independent of the angle~$\beta$.

\begin{figure}
	\centering
	\includegraphics[width=0.9\linewidth]{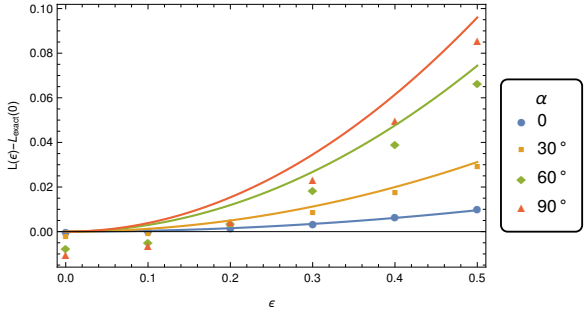}
	\caption{Variation of the spindown luminosity depending on the displacement $\epsilon$ and on obliquity $\alpha$. It is independent of $\beta$ therefore only shown for $\beta=0\degr$.}
	\label{fig:luminosite_a30_depl}
\end{figure}

We compare these analytical approximations to the results of our numerical simulations where we also found that the spindown does not depend on the angle~$\beta$ as expected from eq.~(\ref{eq:Lm1e2}). We performed a set of runs with relevant geometric parameters by varying the set $(\alpha, \beta, \delta, \epsilon)$ and choosing different rotation periods symbolized by the adimensionalized parameter~$a$. For $a=0.2$ we summarize the simulation outputs for $\alpha=\{0\degr, 30\degr, 60\degr, 90\degr\}$ in Fig.~\ref{fig:luminosite_a30_depl}. The luminosity computed from the simulation $L(\epsilon)$ is compared to the exact vacuum centred dipole $L_{\rm exact} = L_{\rm dip} \, \sin^2\alpha$. The corresponding analytical expectations are shown by solid color curves. The agreement is good, the $\epsilon^2$ dependence is clearly retrieved and follows the formula given in eq.~(\ref{eq:Lm1e2}) even for a large off-centred distance with $\epsilon=0.5$. For such a high displacement the quadrupolar expansion is not sufficient to recover the field line topology close to the star but it suffices to compute accurately the spindown luminosity because higher order contributions to the spindown luminosities scale at least as $a^3\,\epsilon^3$. Therefore our approximate expression detailed in eq.~(\ref{eq:Lm1e2}) remains valid even for large displacements~$\epsilon\lesssim1$ as long as the spindown is concerned.

\section{Electromagnetic kick}
\label{sec:Kick}

\begin{figure}
	\centering
	\includegraphics[width=0.9\linewidth]{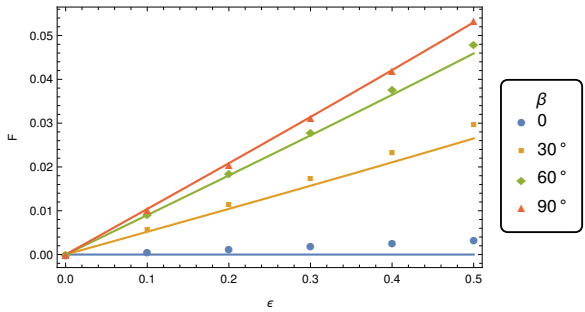}
	\caption{Electromagnetic force induced by a rotating off-centred dipole for different displacements~$\epsilon$ and different angles~$\beta$ for $\alpha=30\degr$.}
	\label{fig:force_a30_depl}
\end{figure}

\begin{figure}
	\centering
	\includegraphics[width=0.9\linewidth]{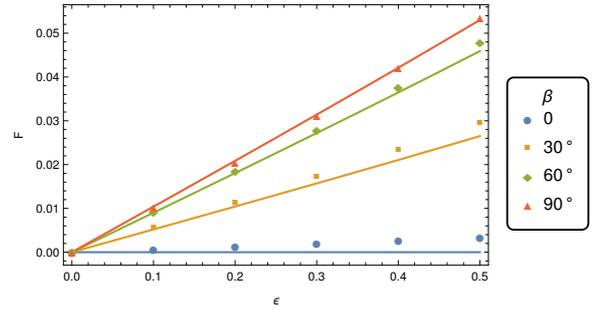}
	\caption{Electromagnetic force induced by a rotating off-centred dipole for different displacements~$\epsilon$ and different angles~$\beta$ for $\alpha=60\degr$.}
	\label{fig:force_a60_depl}
\end{figure}

In \cite{petri_radiation_2016} we gave approximate expressions for the electromagnetic kick felt by the neutron star when radiating as an off-centred dipole. The dipolar $m=1$ and quadrupolar $m=2$ contributions are given respectively by
\begin{subequations}
	\begin{align}
	F_{m=1} & = \frac{6}{5} \, \frac{L_{\rm dip}}{c} \, a\,\epsilon \, \cos\alpha \, \sin\alpha \, \sin\beta \\
	F_{m=2} & = \frac{256}{105} \, \frac{L_{\rm dip}}{c} \, a^3\,\epsilon^3 \, \cos\alpha \, \sin\alpha \, \sin\beta  .
	\end{align}
\end{subequations}
These formula are again only valid for $\delta=90\degr$. The quadrupolar term remains negligible and a factor $a^2\,\epsilon^2$ weaker than the dipolar term. It can be dropped from the estimate without introducing much error in the estimate of the electromagnetic force. Indeed, we computed this force from our set of runs presented in the previous section. A compilation of the results is shown for $\alpha=30\degr$ in Fig.~\ref{fig:force_a30_depl} and for $\alpha=60\degr$ in Fig.~\ref{fig:force_a60_depl}. The linear scaling with respect to~$\epsilon$ is retrieved to good accuracy as seen from the color solid lines representing the analytical expectations. 
Actually, both plots are very similar because they only differ by the expression $\cos\alpha\,\sin\alpha$ which are numerically identical if $\alpha=30\degr$ or $\alpha=60\degr$.

\begin{figure}
	\centering
	\includegraphics[width=0.9\linewidth]{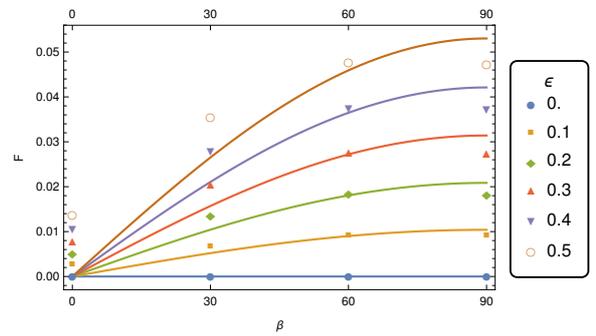}
	\caption{Electromagnetic force induced by a rotating off-centred dipole for different displacements~$\epsilon$ and different angles~$\beta$ for $\alpha=30\degr$ showing the $\sin\beta$ dependence.}
	\label{fig:force_a30_beta}
\end{figure}

\begin{figure}
	\centering
	\includegraphics[width=0.9\linewidth]{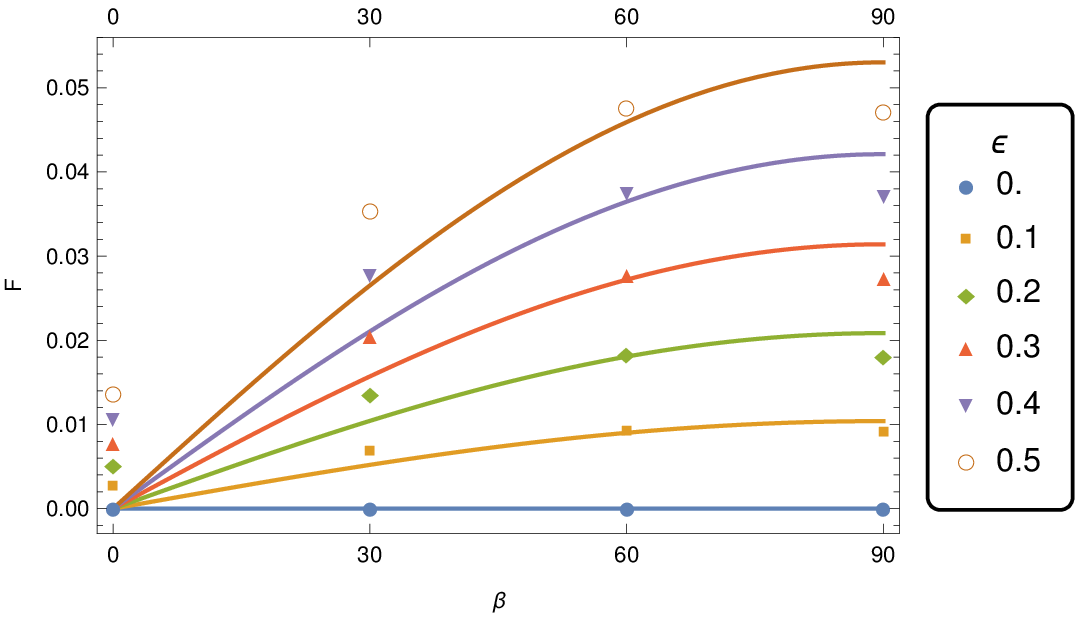}
	\caption{Electromagnetic force induced by a rotating off-centred dipole for different displacements~$\epsilon$ and different angles~$\beta$ for $\alpha=60\degr$ showing the $\sin\beta$ dependence.}
	\label{fig:force_a60_beta}
\end{figure}

Next we show the dependence on the angle~$\beta$ for $\alpha=30\degr$ in Fig.\ref{fig:force_a30_beta} and for $\alpha=60\degr$ in Fig.\ref{fig:force_a60_beta}. The accuracy with only the dipolar term is already remarkable, pointing out the $\sin\beta$ dependence. Only the $\beta=0\degr$ geometry is not well reproduced. Again, both kicks are hardly distinguishable because of the $\cos\alpha\,\sin\alpha$ dependence and the particular value of $\alpha$ chosen.

We finish this theoretical study by the electromagnetic torque calculation.

\section{Electromagnetic torque}
\label{sec:Couple}

Following \cite{beskin_anomalous_2014}, we compute the electromagnetic torque exerted on the neutron star surface according to the Laplace force given by
\begin{equation}
\label{eq:Couple_Laplace}
\mathbf K = R^3 \, \iint \mathbf [ \sigma_{\rm s} \, \mathbf n \wedge \mathbf E + ( \mathbf B \cdot \mathbf n ) \, \mathbf i_{\rm s} ] \, d\Omega 
\end{equation}
where $\mathbf n$ is the unit normal to the surface, $\sigma_{\rm s} = \varepsilon_0 \, [\mathbf{E}] \cdot \mathbf{n}$ represents the surface charge density and $\mu_0 \, \mathbf i_{\rm s} = \mathbf{n} \wedge [\mathbf{B}]$ the surface current density. The notation $[\mathbf{F}]$ means the jump of the vector field $\mathbf{F}$ across the layer. Because of the perfect conductor assumption inside the star, there is no volume contribution to the torque as by definition $\rho \, \mathbf{E} + \mathbf{j} \wedge \mathbf{B} = \mathbf{0}$ where $(\mathbf{E},\mathbf{B})$ is the electromagnetic field, $\rho$ the charge density and $\mathbf{j}$ the current density inside the star. $\varepsilon_0$ and $\mu_0$ are the vacuum permittivity and permeability.

Expressions for the torque are lengthy for a decentred dipole. We only give the approximation for $\delta=90\degr$. In this special case, the components of the electric and magnetic torques, respectively denoted by $K^{\rm E}$ and $K^{\rm B}$, are given to lowest order in $a=R/\rlight$ by a decomposition onto the orthonormal basis $(\ex, \ey, \ez)$ with components
\begin{subequations}
	\label{eq:Torque}
\begin{align}
 K_{\rm x}^{\rm E} & = \frac{L_{\rm dip} \, \sin 2 \alpha}{105000 a\,\Omega}  (350 a^3 (a^2 (162 \epsilon ^2+5)+162 	\epsilon ^2) \cos \beta - \nonumber \\
 & 3 (a^4 (3744 \epsilon ^2+875)+a^2 (1750-21870 \epsilon ^2)+100 (81 \epsilon ^2+140)) \sin \beta ) \\
 K_{\rm y}^{\rm E} & = \frac{L_{\rm dip} \, \sin 2 \alpha}{105000 a\,\Omega}  (3 (a^4 (3552 \epsilon ^2+875) - 70 a^2 	(468 \epsilon ^2-25) \nonumber \\
 & +200 (27 \epsilon ^2+70)) \cos 	\beta +350 a^3 (a^2 (243 \epsilon ^2+5)+243 \epsilon ^2)
 	\sin \beta ) \\
 K_{\rm x}^{\rm B} & = \frac{L_{\rm dip} \, \sin 2 \alpha}{70000 a\,\Omega}  (1400 a (a^2 (93 \epsilon ^2-25)+25) 	\cos \beta  \nonumber \\
 & +(3 a^4 (75228 \epsilon ^2-11375) \nonumber \\
 & -1470 a^2 (66 	\epsilon ^2-25)-100 (324 \epsilon ^2+245)) \sin \beta ) \\
 K_{\rm y}^{\rm B} & = -\frac{L_{\rm dip} \, \sin 2 \alpha}{70000 a\Omega}  ((a^4 (222056 \epsilon ^2-34125)+a^2 	(36750-83280 \epsilon ^2) \nonumber \\
 & -100 (216 \epsilon ^2+245)) 	\cos \beta \nonumber \\
  & -700 a (a^2 (207 \epsilon ^2-50)+50) \sin \beta
 	) \\
 K_{\rm z}^{\rm B} & = \frac{L_{\rm dip}}{50\,\Omega} \, ((a^2 (216 \epsilon ^2-25)+25) \cos 2 \alpha +a^2 (25-264 \epsilon ^2)-25)
\end{align}
\end{subequations}
The electric torque along the $z$ axis always vanishes $K_{\rm z}^{\rm E} =0$.

\begin{figure}
	\centering
	\includegraphics[width=0.9\linewidth]{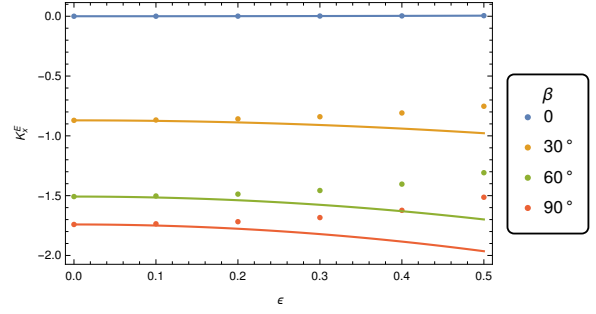}
	\caption{$K_{\rm x}^{\rm E}$ component of the electric torque induced by a rotating off-centred dipole for different displacements~$\epsilon$ and angles~$\beta$ for $\alpha=60\degr$.}
	\label{fig:couple_E_x_r0.2_a60_beta}
\end{figure}

\begin{figure}
	\centering
	\includegraphics[width=0.9\linewidth]{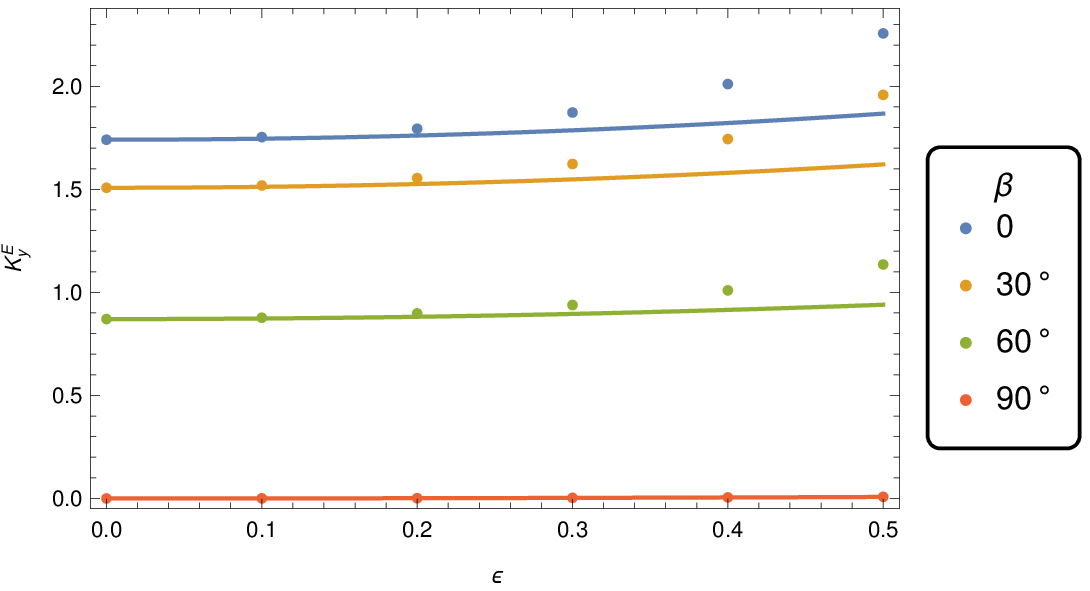}
	\caption{$K_{\rm y}^{\rm E}$ component of the electric torque induced by a rotating off-centred dipole for different displacements~$\epsilon$ and angles~$\beta$ for $\alpha=60\degr$.}
	\label{fig:couple_E_y_r0.2_a60_beta}
\end{figure}

\begin{figure}
	\centering
	\includegraphics[width=0.9\linewidth]{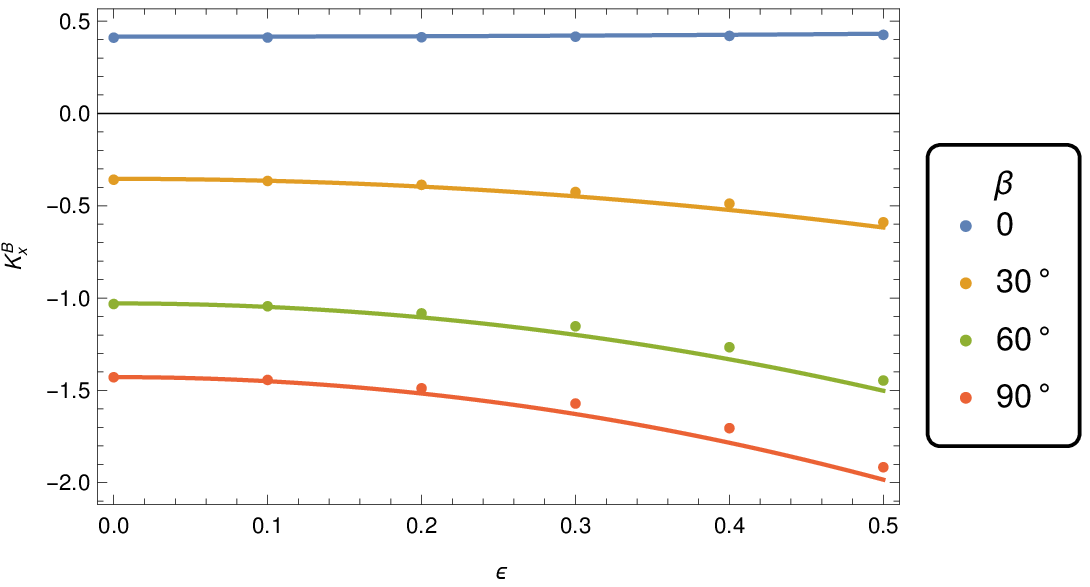}
	\caption{$K_{\rm x}^{\rm B}$ component of the magnetic torque induced by a rotating off-centred dipole for different displacements~$\epsilon$ and angles~$\beta$ for $\alpha=60\degr$.}
	\label{fig:couple_B_x_r0.2_a60_beta}
\end{figure}

\begin{figure}
	\centering
	\includegraphics[width=0.9\linewidth]{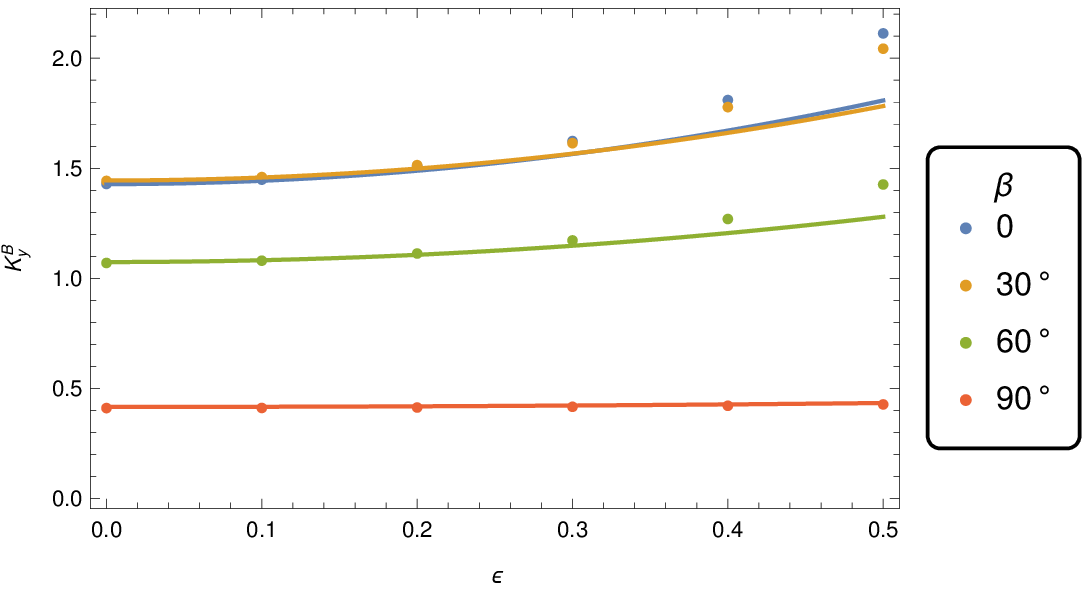}
	\caption{$K_{\rm y}^{\rm B}$ component of the magnetic torque induced by a rotating off-centred dipole for different displacements~$\epsilon$ and angles~$\beta$ for $\alpha=60\degr$.}
	\label{fig:couple_B_y_r0.2_a60_beta}
\end{figure}

\begin{figure}
	\centering
	\includegraphics[width=0.9\linewidth]{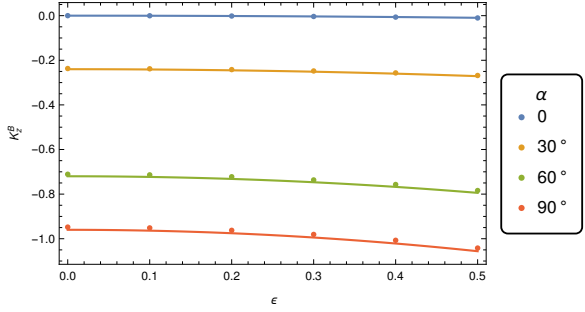}
	\caption{$K_{\rm z}^{\rm B}$ component of the magnetic torque induced by a rotating off-centred dipole for different displacements~$\epsilon$ and angles~$\alpha$. This component is independent of $\beta$.}
	\label{fig:couple_B_z_r0.2_a60_beta}
\end{figure}

We compare these analytical expressions~(\ref{eq:Torque}) to the torque found by integration of eq.~(\ref{eq:Couple_Laplace}) directly from the simulations. The results are shown individually for the electric torque $K_{\rm x}^{\rm E}$ along the $x$ axis in Fig.~\ref{fig:couple_E_x_r0.2_a60_beta}, the electric torque $K_{\rm y}^{\rm E}$ along the $y$ axis in Fig.~\ref{fig:couple_E_y_r0.2_a60_beta}, the magnetic torque $K_{\rm x}^{\rm B}$ along the $x$ axis in Fig.~\ref{fig:couple_B_x_r0.2_a60_beta}, the magnetic torque $K_{\rm y}^{\rm B}$ along the $y$ axis in Fig.~\ref{fig:couple_B_y_r0.2_a60_beta} and magnetic torque $K_{\rm z}^{\rm B}$ along the $z$ axis in Fig.~\ref{fig:couple_B_z_r0.2_a60_beta}. To a good accuracy, the simulation results agree with the analytical approximations even for high displacements~$\epsilon\approx0.5$.

After this rather theoretical investigation of the off-centred rotating dipole, showing the induced spindown luminosity perturbation, the kick and additional torque imparted to the star, we finish this paper by a last section about possible consequences for neutron stars in binary systems.

\section{Impact on binary neutron star systems}
\label{sec:Binaire}

Several hundreds of neutron star binary systems are known to date. The usual channels to form such binaries are well described in several papers like \cite{phinney_pulsars_1992} or \cite{postnov_evolution_2014}. Moreover, \cite{tauris_formation_2017} give an excellent review about the more specific double neutron star systems, pointing out their characteristics, formation and evolution. Neutron star binaries are expected to relax to almost circular orbits with very low eccentricities~$e\approx 0$ due to mass transfer and tidal circularization. Large eccentricities $e\gtrsim0.3$ can however be produced by a supernova explosion in double neutron stars when a significant fraction of the binary mass is lost. Here surprisingly, some double neutron stars show low eccentricities requiring alternative binary evolution scenarios. In this last section, we show how the electromagnetic kick produced by an off-centred magnetic dipole can modify the orbital eccentricity, sometimes generating moderate to large eccentricities in neutron star binaries.

Let us assume that both stars, labelled with subscript $1$ and $2$ are subject to an electromagnetic kick respectively denoted by $\mathbf{F}_1$ and $\mathbf{F}_2$. The equations of motion in the inertial observer frame for this problem are given by
\begin{subequations}
	\begin{align}
	\ddot{\mathbf{r}}_1 & = G \, m_2 \, \frac{\mathbf{r}}{r^3} + \frac{\mathbf{F}_1}{m_1} \\
	\ddot{\mathbf{r}}_2 & = - G \, m_1 \, \frac{\mathbf{r}}{r^3} + \frac{\mathbf{F}_2}{m_2}
	\end{align}
\end{subequations}
where $\mathbf{r}= \mathbf{r}_2 - \mathbf{r}_1$ and $\mathbf{r}_i$ is the position of star~$i$ and dots meaning time derivation. The barycenter, defined by
\begin{equation}\label{eq:Barycentre}
\mathbf{R} = \frac{m_1 \, \mathbf{r}_1 + m_2 \, \mathbf{r}_2}{m_1+m_2}
\end{equation}
feels a net force according to Newton second law
\begin{equation}
\label{eq:barycentre}
\ddot{\mathbf{R}} = \frac{\mathbf{F}_1+\mathbf{F}_2}{m_1+m_2} .
\end{equation}
It therefore accelerates at a constant rate. In order to remove this uniform acceleration not relevant for the orbital parameters evolution, we express the motion in the barycenter frame by introducing two new position vectors
\begin{subequations}
	\begin{align}
	\mathbf{\brho}_1 & = \mathbf{r}_1 - \mathbf{R} \\
	\mathbf{\brho}_2 & = \mathbf{r}_2 - \mathbf{R} .
	\end{align}
\end{subequations}
These vectors satisfy the equations of motion
\begin{subequations}
	\begin{align}
	\ddot{\mathbf{\brho}}_1 & = G \, m_2 \, \frac{\mathbf{r}}{r^3} + \frac{1}{m_1+m_2} \left( \frac{m_2}{m_1} \, \mathbf{F}_1 - \mathbf{F}_2 \right)\\
	\ddot{\mathbf{\brho}}_2 & = - G \, m_1 \, \frac{\mathbf{r}}{r^3} + \frac{1}{m_1+m_2} \left( \frac{m_1}{m_2} \, \mathbf{F}_2 - \mathbf{F}_1 \right) .
	\end{align}
\end{subequations}
The equivalent one body problem then reads
\begin{equation}
\label{eq:AcceleratedKeplerProblem}
\ddot{\mathbf{r}} = - G \, (m_1+m_2) \, \frac{\mathbf{r}}{r^3} + \frac{\mathbf{F}_2}{m_2} - \frac{\mathbf{F}_1}{m_1} .
\end{equation}
This tow-body problem in gravitation, subject to an additional acceleration expressed by
\begin{equation}
\label{eq:Acceleration}
 \mathbf{A} = \frac{\mathbf{F}_2}{m_2} - \frac{\mathbf{F}_1}{m_1}
\end{equation}
arising from an external body, an external pressure or produced by the stars themselves, is known as the Stark or accelerated Kepler problem \citep{namouni_accelerated_2007}. Moreover, if this force is constant in direction and time, interestingly enough, the problem is fully integrable as shown in depth by \cite{lantoine_complete_2011}. They give a complete set of analytical solutions involving parabolic coordinates and elliptic functions for elliptic, parabolic and hyperbolic trajectories. In the present work, we focus only on bound orbits for a constant in direction acceleration~$\mathbf{A}$. Important but simple results for eccentricity excitation have been reported by \cite{namouni_origin_2005} for extrasolar planets and for jets by \cite{namouni_accelerated_2007}. In all these cases, the accelerating field is derived from the gradient of a scalar function. Therefore, the time evolution of the orbital parameters are known to satisfy Lagrange planetary equations. A detailed derivation of the formalism and practical results can be found in \cite{beutler_methods_2005}. Next we discuss a straightforward application to binaries in circular orbit for which simple analytical expressions have been derived.

Indeed, let us describe the time dependence of the eccentricity~$e(t)$ when starting from a circular orbits with $e(t=0)=0$. It varies periodically following a sinusoidal law expressed as
\begin{equation}
\label{eq:EvolutionEccentricite}
e(t) = \left| \sin i_0 \, \sin\left( \frac{3\,A}{2\,\Omega\,a} \, t\right) \right|
\end{equation}
where $a$ is the semi-major axis, $A=\mathbf{\|A\|}$ is the acceleration produced by the neutron stars themselves due to the electromagnetic kick~$F_i$, $i_0$ is the inclination angle between the acceleration vector~$\mathbf{A}$ and the orbital angular momentum vector and $\Omega = \sqrt{G \, (m_1+m_2)/a^3}$ is the keplerian frequency. The typical timescale of eccentricity excitation is then
\begin{equation}
\label{eq:TempsExcentricite}
T_e = \frac{\upi\,\Omega\,a}{3\,A} =  \frac{\upi}{3\,A} \, \sqrt{\frac{G \, (m_1+m_2)}{a}}.
\end{equation}
Note the factor~$2$ difference with respect to the $\sin$ argument because of the absolute sign, dividing the period by the same factor~2. This order of magnitude is easily reproduced by noting that generating a significant eccentricity requires the accelerating field~$A$ to produce orbital velocities of the order of the keplerian velocity $v=\Omega\,a$, thus equating $A\,T_e = v$ which agrees with Eq.~(\ref{eq:TempsExcentricite}) within a factor unity. This timescale holds if the excitation~$A$ is constant in time. However, the acceleration~$\mathbf{A}$ depends on the electromagnetic kick~$\mathbf{F}$ which itself depends linearly on the spindown luminosity~$L$ because $F \propto L/c$, see previous sections. For a magnetic dipole rotating in vacuum, a braking index of $n=3$ (which still hold to good accuracy for an off-centred dipole) leads to a decreasing luminosity according to
\begin{equation}\label{eq:SpindownTemps}
L(t) = \frac{L_0}{(1+t/\tau_{\rm c})^2}
\end{equation}
where $\tau_{\rm c}=P/2\dot{P}$ is the characteristic electromagnetic spindown time scale, the characteristic age of the pulsar. The eccentricity then follows from 
\begin{equation}
\label{eq:EvolutionEccentriciteTemps}
e(t) = \left| \sin i_0 \, \sin\left( \frac{3}{2\,\Omega\,a} \, \int_0^t A(t) \, dt \right) \right| .
\end{equation}
From the spindown decrease in Eq.~(\ref{eq:SpindownTemps}), we easily integrate the action of the acceleration acting on a characteristic time scale~$\tau_{\rm c}$ to find
\begin{equation}
 \int_0^{\tau_{\rm c}} A(t) \, dt = \frac{A(t=0) \, \tau_{\rm c}}{2} .
\end{equation}
After a time equal to the characteristic age, the acceleration becomes negligible. Thus the estimate given for a constant in time acceleration is a good guess if time~$t$ is replaced by $\tau_{\rm c}/2$. The actual binary eccentricity is therefore
\begin{equation}
\label{eq:EccentriciteFinale}
e(\tau_{\rm c}) = \left| \sin i_0 \, \sin\left( \frac{3\,A \, \tau_{\rm c}}{4\,\Omega\,a} \right) \right| .
\end{equation}
Eq.~(\ref{eq:EvolutionEccentricite}) shows that the maximal eccentricity achieved after a full excitation period~$T_e$ is $\sin i_0$. This time scale must be compared to other typical time scales like the true age of the binary and the electromagnetic spindown time scale. It can be shown that the eccentricity depends only on $P$ and $P_{\rm orb}$ but not on $\dot{P}$. Straightforward calculations give
\begin{equation}
e(\tau_{\rm c}) = \left| \sin i_0 \, \sin\left( \frac{9\,\upi^{5/3}}{5\times 2^{1/3}} \, \epsilon \, \frac{I\,P^{-2} \, P_{\rm orb}^{1/3}}{m_1\,c\,\sqrt{G\,(m_1+m_2)}} \right) \right| 
\end{equation}
$m_1$ being the mass of the pulsar and $m_2$ the mass of its companion. For low eccentricities, this expression reduces to
\begin{subequations}
\begin{align}
e(\tau_{\rm c}) & \approx \frac{9\,\upi^{5/3}}{5\times 2^{1/3}} \, \epsilon  \, | \cos\alpha \, \sin\alpha \, \sin\beta \, \sin i_0 | \, \frac{I\,P^{-2} \, P_{\rm orb}^{1/3}}{m_1\,c\,\sqrt{G\,(m_1+m_2)}} \\
& \approx 7.5 \times 10^{-6} \, \epsilon \, | \cos\alpha \, \sin\alpha \, \sin\beta \, \sin i_0 \, | \left( \frac{P}{1~\textrm{s}}\right)^{-2} \, \left(\frac{P_{\rm orb}}{1~\textrm{day}}\right)^{1/3}
\end{align}
\end{subequations}
showing the dependence
\begin{equation}
\label{key}
e \propto \epsilon \, \cos\alpha \, \sin\alpha \, \sin\beta \, P^{-2} \, P_{\rm orb}^{1/3} .
\end{equation}

The eccentricity in Eq.~(\ref{eq:EccentriciteFinale}) can be compared with observations through the ATNF Pulsar Catalogue (\url{http://www.atnf.csiro.au/research/pulsar/psrcat/}) of \cite{manchester_australia_2005}. For concreteness, we set the inclination angle to~$i_0=90\degr$, representing the most favourable geometry to efficiently excite the eccentric orbit. For the accelerating electromagnetic kick, we set an average of $\epsilon \, \cos\alpha \, \sin\alpha \, \sin\beta = 0.01$ and the total mass of the binary to $m_1 + m_2 = 2.5\,M_\odot$. The neutron star radius is 12~km and its moment of inertia is~$I=10^{38}$~kg.m$^2$.

\begin{figure}
	\centering
	\includegraphics[width=0.9\linewidth]{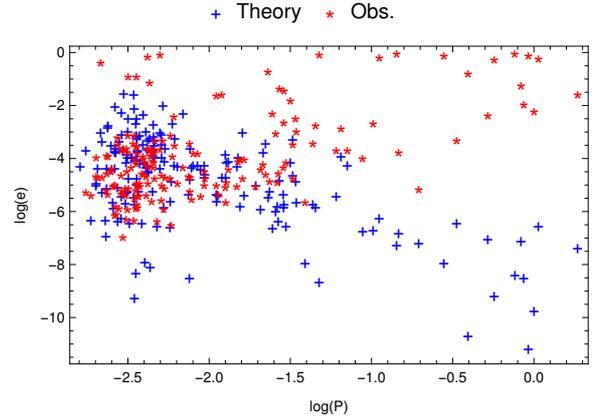}
	\caption{Observed (red) and expected (blue) relation between spin period~$P$ and eccentricity~$e$ for neutron stars binaries.}
	\label{fig:ecc_all_period}
\end{figure}

\begin{figure}
	\centering
	\includegraphics[width=0.9\linewidth]{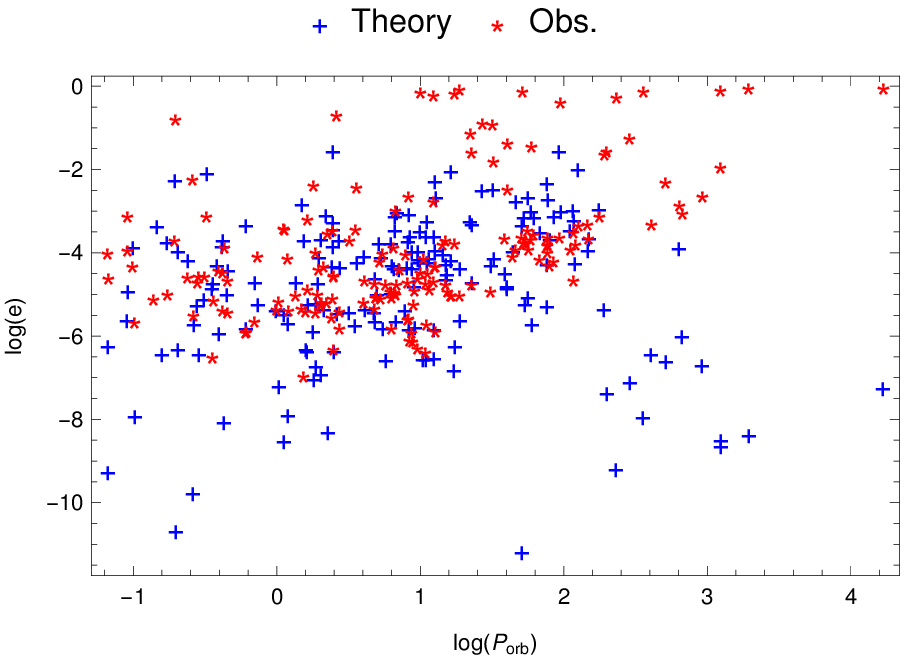}
	\caption{Observed (red) and expected (blue) relation between orbital period~$P_{\rm orb}$ and eccentricity~$e$ for neutron stars binaries.}
	\label{fig:ecc_all_porb}
\end{figure}

\begin{figure}
	\centering
	\includegraphics[width=0.9\linewidth]{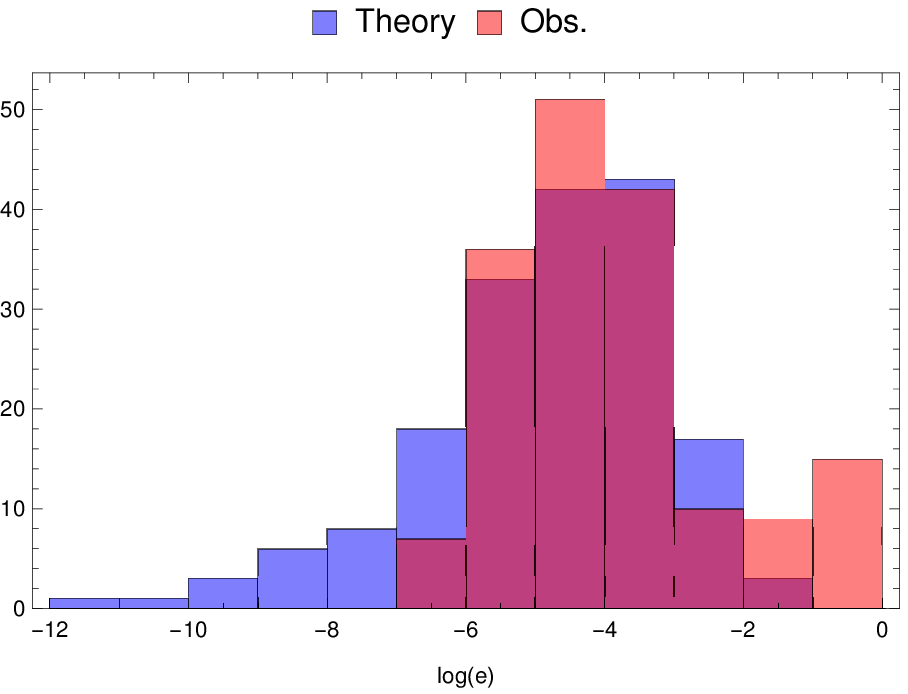}
	\caption{Comparison of the observed (red) and expected (blue) population of eccentricities~$e$ for neutron stars binaries with $\epsilon \, \cos\alpha \, \sin\alpha \, \sin\beta=0.01$.}
	\label{fig:ecc_all_histo}
\end{figure}

\begin{figure}
	\centering
	\includegraphics[width=0.9\linewidth]{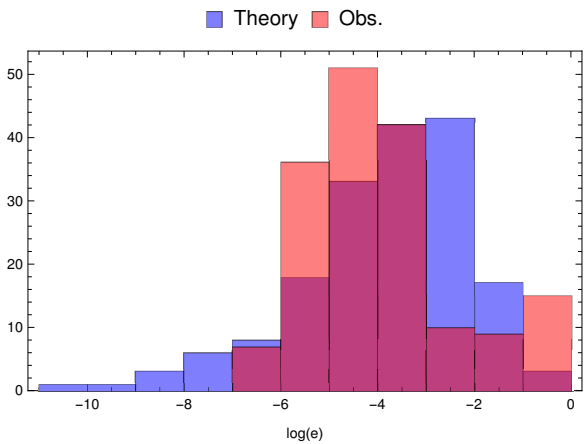}
	\caption{Comparison of the observed (red) and expected (blue) population of eccentricities~$e$ for neutron stars binaries with $\epsilon \, \cos\alpha \, \sin\alpha \, \sin\beta=0.1$.}
	\label{fig:ecc_all_histo-1}
\end{figure}

Fig.~\ref{fig:ecc_all_period} shows the period-eccentricity relation $(P,e)$ for the neutron star binaries contained in the ATNF catalog in red, compared to the model predictions in blue. Pulsars with period less than $P<0.1$~s are reasonably well reproduced with largest eccentricities $e\lesssim0.1$, but longer period pulsar eccentricities predictions are orders of magnitude lower than those observed. There are obviously other means to achieve very large eccentricities~$e\lesssim1$ like for instance a supernova explosion leading to a substantial mass loss in the system. All possible perturbations of orbital parameters must be combined to extract realistic binary geometries. However this is out of the scope of this work that only emphasizes the non negligible role of the electromagnetic kick. Fig.~\ref{fig:ecc_all_porb} compares the observed and predicted orbital period-eccentricity relation $(P_{\rm orb}, e)$. Observed eccentricities for orbital periods less than $P_{\rm orb}<100$~days are well reproduced, but longer orbital period eccentricities here also are orders of magnitude lower than observed. Obviously again, the formation and evolution of the binary system must account for those large eccentricities~$e$. Finally, the eccentricity distribution function is plotted in Fig.~\ref{fig:ecc_all_histo}, showing the range of eccentricities achieved by the electromagnetic kick in blue, starting from a perfect circular orbit and compared to observations in red. The peak arises at $e\approx 10^{-5}-10^{-4}$ with minimum value of $e_{\rm min} \approx 10^{-12}$ and maximum value of $e_{\rm max} \approx 10^{-1}$. The distribution function is well reproduced in the range $[10^{-7},10^{-2}]$. The deficit in high eccentricities for large spin periods and large orbital periods is clearly identified. Amplifying the excitation acceleration~$A$ will produce eccentricities closer to~1. Indeed, when $A$ is increased by a factor 10, i.e. $\epsilon \, \cos\alpha \, \sin\alpha \, \sin\beta = 0.1$, the eccentricity distribution function shown in Fig.~\ref{fig:ecc_all_histo-1} is shifted one order of magnitude to larger $e$ with $e_{\rm max} \gtrsim0.1$. Precise values of the acceleration~$A$ are geometry dependent and only barely constrained. Nevertheless, already small off-centered dipoles with $\epsilon \, \cos\alpha \, \sin\alpha \, \sin\beta \ll 1$ can accounted for the majority of low eccentricity neutron star binaries.

A complete scenario of binary neutron star formation and evolution must account for not only the dramatic birth event but also for secular change in the orbits due to continuous perturbations induced by the stars themselves. We hope that the present study will help to better understand the formation of binary neutron star systems.

\section{Conclusions}
\label{sec:Conclusion}

We performed accurate time-dependent numerical simulations of an off-centred rotating dipole in vacuum, including contributions from high order multipoles, going beyond the quadrupole expansion as done in previous analytical works. As a diagnostic, we showed magnetic field line structures, spin-down luminosities, induced electromagnetic forces and torques. We demonstrated that the analytical approximations taking into account only perturbations up to the magnetic quadrupole are reliable for quickly computing with satisfactory accuracy the geometric dependence of the spin-down luminosity, the electromagnetic kick and torque according to the orientation and location of the magnetic dipole with respect to the centre of the star.

Other models able to produce naturally multipolar components are those incriminating an inhomogeneous magnetization inside the star. It is well known that a uniformly magnetized sphere produces in vacuum a perfectly centred dipole. If the homogeneity is broken, multipolar components will easily arise in vacuum outside this sphere. Therefore, such models offer an interesting alternative to the off-centred dipole, avoiding a singular magnetic field at the location of the magnetic moment, replacing it by a smooth magnetization distribution over the whole stellar volume.

Electromagnetic kicks and torques are important for understanding the secular evolution of isolated neutron stars, their proper motion and precession, but also for binary neutron stars. Indeed, when in a binary neutron star system, the electromagnetic force permanently perturbs the keplerian orbit and could explain the origin of high eccentricities in several of those binaries as shown in the previous section. The distribution of low eccentricities is well reproduced by an electromagnetic kick acting on the typical spin-down time scale. The electromagnetic torque leads to precession of an isolated pulsar that could be observed in the secular evolution of radio pulse profiles and maybe at higher energies like X-rays and gamma-rays. Such investigations however require knowledge about the radiation mechanisms and is left for future work.

An important extension to the present work is the inclusion of pair plasma within the magnetosphere, screening the electric field component parallel to the magnetic field in the force-free regime. Such models will be computed with our pseudo-spectral Maxwell solver in an upcoming paper.

\section*{Acknowledgements}

This work has been published under the framework of the IdEx Unistra and benefits from a funding from the state managed by the French National Research Agency as part of the investments for the future program. This work is also supported by the CEFIPRA grant IFC/F5904-B/2018.








%


\bsp	
\label{lastpage}
\end{document}